\documentstyle[12pt,epsfig]{article}

\begin{document}

\title{Exploring the Gamma Ray Horizon\\
with the next generation of Gamma Ray Telescopes.\\
Part 1: Theoretical predictions.}
\author{O.Blanch, M.Martinez\\
{\it IFAE, Barcelona (Spain) }}

\maketitle

\begin{abstract}
The physics potential of the next generation of Gamma Ray
Telescopes in exploring the Gamma Ray Horizon is discussed. It is
shown that the reduction in the Gamma Ray detection threshold might
open the window to use precise determinations of the Gamma Ray
Horizon as a function of the redshift to either put strong constraints
on the Extragalactic Background Light modeling or to obtain relevant
independent constraints in some fundamental
cosmological parameters.
\end{abstract}


\newpage


\section{Introduction}

Imaging Cherenkov Telescopes have proven to be the most successful
tool developed so far to explore the cosmic gamma rays of energies
above few hundred GeV. A pioneering generation of installations
has been able to detect a handful of sources and start a whole
program of very exciting physics studies. Now a second generation
of more sophisticated Telescopes is starting to operate and is
providing already with new exciting observations. One of the main
characteristics of some of these new Telescopes \cite{MAGIC}, is
the potential ability to reduce the gamma ray energy threshold
below $\sim 10-20$ GeV, helping to fill the existing observational
energy gap between the detector on satellites and the ground-based
installations.\par

In the framework of the Standard Model of particle
interactions, high energy gamma rays traversing cosmological
distances are expected to be absorbed by their interaction with
the diffuse background radiation fields, or ``Extragalactic Background
Light'' (EBL), producing $e^+ e^-$ pairs. The
$ \gamma_{HE} \gamma_{EBL} \rightarrow e^+ e^- $ cross section is
strongly picked to $E_{CM} \sim 1.8 \times (2 m_e c^2)$ and therefore,
there is a specific range in the EBL energy which is ``probed''
by each gamma ray energy \cite{Vassiliev}.\par

This effect should lead to the existence of a ``Gamma Ray Horizon'',
limiting the feasibility of observing very high energy gamma rays
coming from very far distances. The actual value of this horizon
distance for gamma rays of a given energy, depends on the number
density of the diffuse background radiation of the relevant energy
range, which is traversed by the gamma rays. In the range of gamma
ray energies which can be effectively studied by the next generation
of Gamma Ray telescopes (from, say, 10
GeV to 50 TeV), the most relevant EBL component is the ultraviolet
(UV) to infrared (IR) contribution.\par

Several models have been developed to try to predict that EBL
density \cite{COBE-IV,Kneiske-Model}. These models do a quite
complex convolution of the measurements of star formation rate,
initial mass function and dust and light recycling history. The
result is a set of relatively model-independent predictions which
accuracy is improving as the quality of their astrophysics inputs
improves with the new deep-field observations and which fits
reasonably well the existing data.\par

Therefore, quantitative predictions of the Gamma Ray Horizon have already
been made but, unfortunately, so far no clear confirmation can be
drawn from the observations of the present generation of Gamma Ray
Telescopes.\par

On the one hand, some very high energy gamma ray events might have been
observed from Mkn 501, a blazar at redshift $z \sim 0.03$
\cite{Mkn501TeV}. The mere observation of these events would somehow
contradict the above predictions indicating, might be, the presence of a
new mechanism violating the forementionned gamma-gamma reaction threshold, for
which, for instance, Lorenz-Invariance violation has been advocated,
as we'll discuss later.
Unfortunately, the statistics is scarce and for these events the
actual systematic uncertainty in the energy determination might be
large and hence the situation remains somewhat unclear.
On the other hand, for the handful of presently well established
extragalactic sources (all of them at modest redshifts), no clear
observation of a common energy cutoff which could be attributed to
the gamma absorption in the intergalactic medium instead of simply
to internal source characteristics, has been established so far.
Nevertheless, for Mkn 501 a clear exponential energy spectrum cutoff
has been observed and, under the assumption that its origin is the
EBL absorption, upper limits on the EBL density in agreement with
the expectations have been placed \cite{Mkn501EBL}.\par

The fact that the new generation of Cherenkov Telescopes will
reach a considerably lower energy threshold than the previous one
should be of paramount importance in improving the present
experimental situation for, at least, two reasons:

\begin{itemize}
\item{} Lower energy points with a much smaller uncertainty, due to
the steep spectra, will be added to the spectra of the already
observed sources allowing to disentangle much better the overall flux
and spectral index from the cutoff position in the spectrum fit.
\item{} Sources at higher redshift should be observable, giving a
stronger lever arm in constraining the predictions and the possibility
of observing a plethora of new sources that will allow unfolding the
emission spectra and the gamma absorption.
\end{itemize}

The goal of this work is to analyse the physics potential of this
new generation of telescopes in the measurement of the Gamma Ray
Horizon and more specifically its impact in the understanding
of the various models and parameters involved in its predictions.
For this, the work is structured in two parts.\par

The first part, which is the one covered in the present paper,
concerns the theoretical predictions. In this part, first the
definition of the terms used in this work and their calculational
procedure are reviewed in detail. Then, the
theoretical predictions obtained for the Gamma Ray Horizon for
different EBL approaches and also for
different cosmological parameter sets are presented. Finally
plausible scenarios that could effect the Gamma Ray Horizon
predictions are commented. The actual
sensitivity to these models and parameters is discussed.\par

The second part, which will be presented in a fore-coming paper,
will deal with the prospective on what the experimental scenario
might look like and a discussion on how much one can expect to
pinpoint the parameters of the theory (with special emphasis on
the cosmological ones)in the extrapolated data scenario.\par

\section{Description of the calculation}

In this section the detailed calculation of the Gamma Ray Horizon in
terms of the predicted EBL density spectra is presented. Our strategy
has been performing the complete calculation without any approximation by
using a numerical integration approach. Different ansatzs for the
calculation of the EBL predictions have been also analysed to
see their impact on the Gamma Ray Horizon prediction. Also, the
dependence on all the intervening parameters has been kept explicit
to be able to track the effect of the different hypotheses in the
final prediction.\par

The optical depth can be written with its explicit redshift and energy
dependence\cite{Stecker-95} as

\noindent
\begin{equation}
\tau(E,z) =
\int_{0}^{z}dz'\frac{dl}{dz'}\int_{0}^{2}dx \,
\frac{x}{2}\int_{\frac{2m^{2}c^{4}}{Ex(1+z')^{2}}}^{\infty}
d\epsilon\cdot n(\epsilon,z') \cdot \sigma[2xE\epsilon(1+z')^{2}]
\label{eq:OpD}
\end{equation}

where $x \equiv 1-\cos\theta $, $E$ is the energy of the
$\gamma$-ray,$\epsilon$ is the energy of the EBL photon, z is the redshift
of the considered source and $n( \epsilon ,z')$ is the spectral
density at the given z'.\par

The predicted value of the optical depth depends on several physical
parameters. A part from the dependence on the actual absorption
process, which enters through the gamma-gamma cross section,
and the direct dependence on the cosmological parameters
$H_{0}$, $\Omega_{M}$,
and $\Omega_{\Lambda}$
introduced by the geodesic radial displacement function, the
spectral energy density is also an input
parameter.\par

There exists observational data with determinations and bounds of the
background energy density at $z=0$ for several energies
\cite{EBLmeasurements}. The determinations come from direct measurements of the
EBL density using instruments on satellites whereas the bounds, happen
mostly in the infrared part of the EBL and come from extrapolations
using galaxy counting. Given the difficulty of observing ``cold
galaxies'' due to the zodiacal light background, they provide just lower
limits.\par

\begin{figure}[h]
  \begin{center}
    \epsfig{file=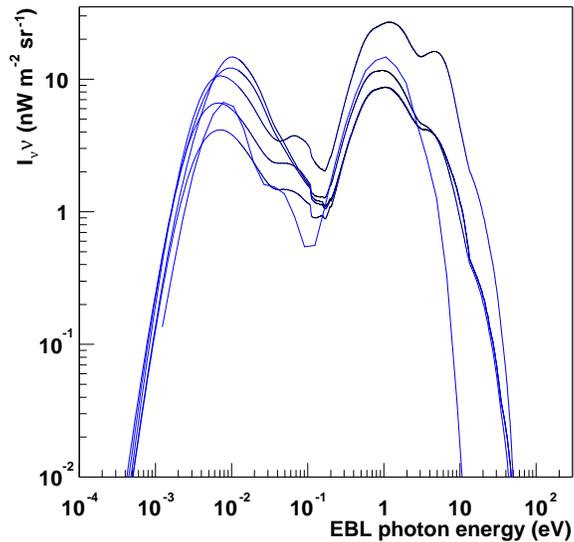,width=8cm}
    \caption{Model predictions for the energy density spectra at $z=0$.}
  \end{center}
    \label{fig:sedatz0}
\end{figure}

Several models, which fit the observational data of $n( \epsilon ,z=0) $,
have been suggested \cite{COBE-IV} (a set of predictions for the most
significant models can be seen in figure~\ref{fig:sedatz0}). These
models do not provide all the necessary information for our
calculation: they provide a description of spectral density at $z=0$
while we need to
know also the evolution of $n( \epsilon )$ as a function of $z$.

In this note three different approaches, which represent somehow
limiting cases in the complexity of the ansatz assumed, for the $z$
evolution of the EBL have been used. Their comparison should give a
feeling on how much the predictions change with the complexity of
the theoretical assumptions
and, hence, they might provide a tool to estimate how large the
theoretical uncertainties in the final predictions might be.
Ordered in increasing complexity, these approaches are:

\begin{quote}
\begin{enumerate}
\item Burst of star formation at high redshift \cite{Mannheim}.
\item Parameterisation of the measured star formation rate \cite{Mannheim}.
\item Star formation rate and star evolution \cite{Kneiske-StarEvo}.
\end{enumerate}
\end{quote}

Computing the GRH for these extreme scenarios allows us to get an
estimation of its maximum uncertainty due to the unprecise EBL knowledge.\par

\subsection{Gamma Ray Horizon}

For any
given gamma ray energy, the Gamma Ray Horizon is defined as the
source redshift for which the optical depth is $\tau(E,z) = 1$. Therefore,
the Gamma Ray Horizon gives, for each gamma ray energy,
the redshift location $z$ of a source for which the intrinsic gamma flux
suffers an e-fold decrease when observed on Earth $z=0$ due to the
gamma-gamma absorption.\par

 In practice, the cut-off due to the Optical Depth is completely
folded with the spectral emission of the gamma source. But on the
other hand, the suppression factor in the gamma flux due to the
Optical Depth depends only (assuming a specific cosmology and spectral
EBL density) on the gamma energy and the redshift of the
source. Therefore, a common gamma energy spectrum behaviour of
a set of different gamma sources at the same redshift is most
likely due to the Optical Depth.\par

The goal of this note has been studying the effect of the cosmological
parameters and the different spectral density models in the Gamma Ray
Horizon predictions for the gamma ray energy region covered by the
next generation of Gamma Ray Telescopes. The results of this study are
presented in the next section.\par

\section{Results}
\subsection{Optical Depth and Gamma Ray Horizon}
\label{sec-Res_OD}

As already mentioned, for any given energy of the gamma ray
that travels through the universe, the probability of interaction
with the EBL photons to create $e^{+}e^{-}$ pairs has a
strong dependence with the energy of the background photons. Roughly
speaking, each gamma energy
``probes'' a different EBL photon energy and therefore, the trends
of the EBL spectrum as a function of the photon energy $\epsilon$
as well as its redshift evolution
are reflected in the Optical Depth as a function of gamma energy $E$.\par

\begin{figure}[h]
  \begin{center}
    \epsfig{file=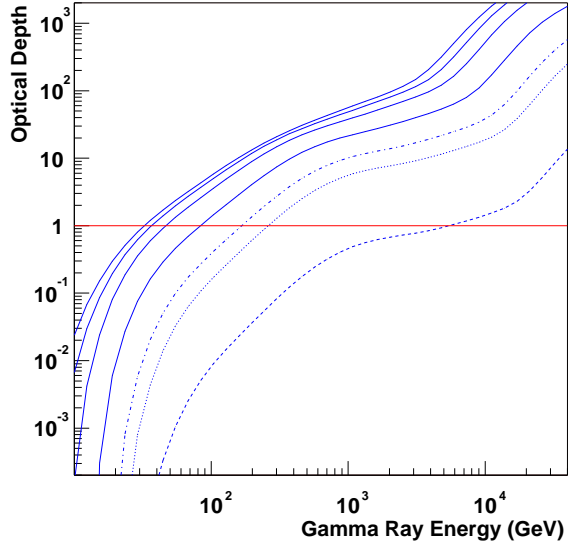,width=8cm}
  \end{center}
    \caption{Optical depth for z = 0.03 (doted line), z = 0.3
    (dashed line), z = 0.5 (dot-dashed line) and z=1,2,3,4 (solid
    lines). The intersection with the horizontal line (Optical Depth $=
    1$) is the Gamma Ray Horizon.}
    \label{fig:OptDep}
\end{figure}

In figure~\ref{fig:OptDep} the Optical Depth for gamma rays coming
from a set of different redshifts are shown as a function of the
gamma ray energy. As already mentioned, in the comoving frame, the
$ \gamma_{HE} \gamma_{EBL} \rightarrow e^+ e^- $ reaction has the
maximum probability when $E_{CM} = E \epsilon (1-\cos\theta) \sim
1.8 \times (2 m_e c^2)$. This means that the flat zone seen in
figure~\ref{fig:OptDep} corresponds to gamma rays that interact
mainly with EBL between roughly $0.2$ eV and $1$ eV (depending on
the source redshift), where the density of EBL photons has a sharp
break down (figure~\ref{fig:sedatz0}). On the other hand, while
the gamma rays explore the peaks due to the star radiation and the
absorption and reemission in the interstellar medium, the Optical
Depth keeps increasing but with a non-constant slope.\par

In figure~\ref{fig:GRHdensity} (dotted line) the GRH that we get solving numerically the equation
$\tau=1$ is shown. On the one hand, it is clear that from redshift
$z=1$ onwards, it is quite flat, so that gammas of energy about
$< 30$ GeV could reach the Earth from any distance in the observable
universe. On the other hand
the GRH depends strongly on the redshift for $z<1$.\par

\subsection{Spectral density}

\begin{figure*}[h]
    \begin{center}
    \mbox{
    \epsfig{file=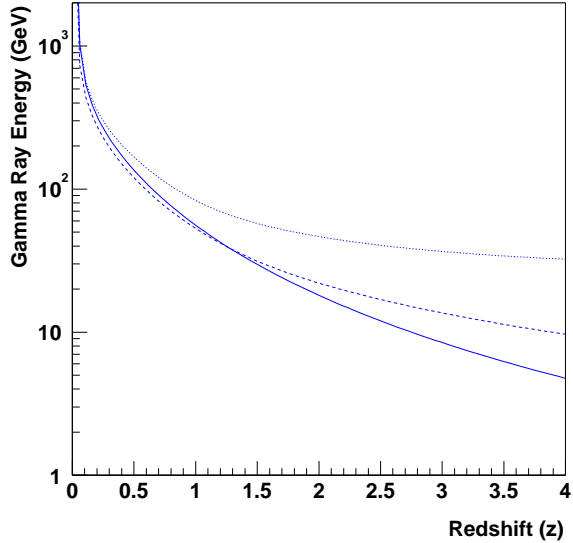,width=8cm} }
    \end{center}
    \caption[]
    {Gamma Ray Horizon for different approaches for the
    calculation of the $z$ evolution of the EBL (see text):
    ``burst of star formation'' (solid
    line), ``star formation rate'' (dashed line), and ``star evolution''
    (doted line).The cosmological parameters are fixed to $H_{0}=68$
    Km s$^{-1}$ Mpc$^{-1}$, $\Omega_{M}=0.35$ and
    $\Omega_{\Lambda}=0.65$.}
    \label{fig:GRHdensity}
\end{figure*}

The Gamma Ray Horizon has been calculated for the three different
evolutions of $n(\epsilon,z')$ already mentioned.  For the first and
second approaches (``Burst of star formation at high redshift'' and
``Parameterisation of the star formation rate'') a model which
defines the $n(\epsilon,0)$ has to be chosen. In
figure~\ref{fig:GRHdensity}, a specific model for $n(\epsilon,0)$
\cite{COBE-IV} and specific values for the Madau curve ($\alpha_{M} =
3.8$, $\beta_{M} = -1$, $z_{b} = 1.5$ and $z_{f} =  10$), which agree
with recent data \cite{Bunker-04}, have been used.\par

The third model, namely the ``Star formation rate and star evolution''
assumption, is likely
the closest to reality, so it
is going to be used for all further studies in this work and, in fact,
has already been used as a
particular example in the previous sections.\par

The fact that we will stick to this approach and that no error bars
are shown in figure~\ref{fig:GRHdensity}, does not mean that this
prediction is free from theoretical uncertainties. This model has a
lot of inputs that come from cosmological measurements which have, in
fact, quite large uncertainties \cite{Kneiske-Model}. For instance, if
one would assume a fit to the star formation rate
in the redshift region for $z>z_{b}$ as the classical Madau curve
\cite{Madau} instead of a slowly decreasing rate,
this would produce a sizeable change in the GRH ($20-40$ GeV)
prediction at large redshift. The uncertainty for low redshifts can be
estimated by computing the GRH for several models of  $n(\epsilon,0)$,
which produce a factor $\sim 5$ difference in the GRH energy prediction for $z<<1$ independently of the model used for its evolution.\par

\subsection{Cosmological parameters}

As we have already commented, some fundamental cosmological parameters
such as the Hubble constant and the cosmological densities play also
an important role in the calculation of the Gamma Ray Horizon since
they provide the bulk of the $z$ dependence of our predictions.\par

Over the last few years, the confidence in the experimental
determinations of these cosmological parameters has increased
dramatically. To understand the effect of moving these parameters,
the values quoted in table~\ref{tab:Bestfit} have been used.\par

\begin{table}[h]
  \begin{center}
    \footnotesize
    \begin{tabular}{|l|l|r|}    \hline
      Parameter   &   Allowed range \\ \hline
      $H_{0}$       &   72$\pm$4        \\
      $\Omega_{\Lambda}$  &   0.72$\pm$0.09    \\
      $\Omega_{M}$  &   0.29$\pm$0.07    \\ \hline
    \end{tabular}
  \end{center}
  \caption[Best current fit values for cosmological parameters.]
    {Best current fit values for
     cosmological parameters with 1 $\sigma$ confidence level\cite{Spergel-03,Wang-03}.}
  \label{tab:Bestfit}
\end{table}

\begin{figure*}[h]
    \begin{center}
    \mbox{
    \epsfig{file=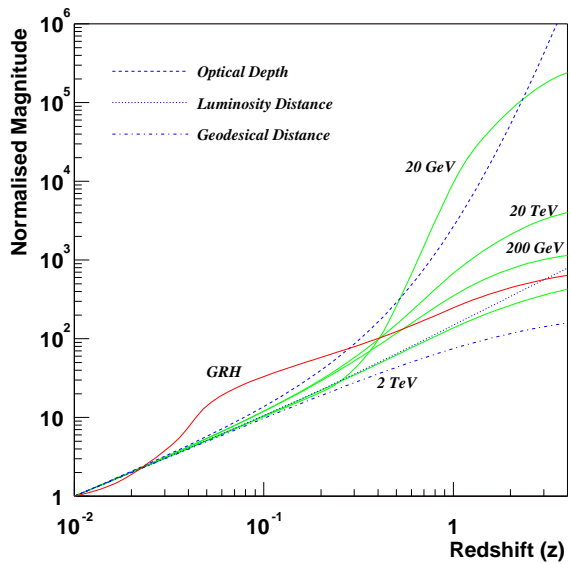,width=8cm}
    }
    \end{center}
    \caption[Redshift dependence of different cosmological observables]
    {Redshift dependence of different observables. The
    predictions are normalized to their value at $z=0.01$.
    The solid lines correspond to the Optical Depth prediction for
    gamma rays of different energies ($20$ GeV to $20$ TeV) while
    the dashed line is the prediction for a flat $\nu I_{\nu}$ EBL
    spectrum. The GRH curve
    gives the $z$ dependence of the inverse of the GRH energy.}
    \label{fig:z-evolution}
\end{figure*}

Before we discuss the impact of each one of these parameters in
our predictions we would like to see how actually the observables
that we will measure (Optical Depths and GRH) depend on the
redshift $z$ to compare it with the redshift dependence of other
observables. For that, we have plotted the prediction for their
$z$ evolution in figure \ref{fig:z-evolution}. In that figure, for
each observable it is shown the prediction normalized to the value
at $z=0.01$. For comparison, the $z$ variation of the
Luminosity-Distance, used for the determination of the
cosmological parameters using Supernova 1A observations and of the
Geodesical-Distance, giving the gamma ray path length, are shown.
One can see that the Optical Depth has a quite different behaviour
depending on the gamma ray energy explored. The $z$ dependence is
very pronounced at large redshifts for $20$ GeV gammas and
approaches a ``Geodesical-Distance''-like shape for $2$ TeV gamma
rays. The reason for that is the actual shape of the EBL spectrum
and its redshift evolution. To give a feeling of the actual
average $z$ dependence of the Optical Depth, the prediction for a
flat $\nu I_{\nu}$ EBL spectrum is also shown. Finally, the $z$
dependence of the inverse of the GRH energy is also shown.\par

Now, to understand up to which level the measurement of the GRH would allow to
get information on $H_{0}$, $\Omega_{\Lambda}$ and $\Omega_{M}$, the
actual prediction of the GRH with the most sophisticated EBL approach
has been repeated for a set of different values of these cosmological
parameters.\par

\begin{figure}[h]
  \begin{center}
    \epsfig{file=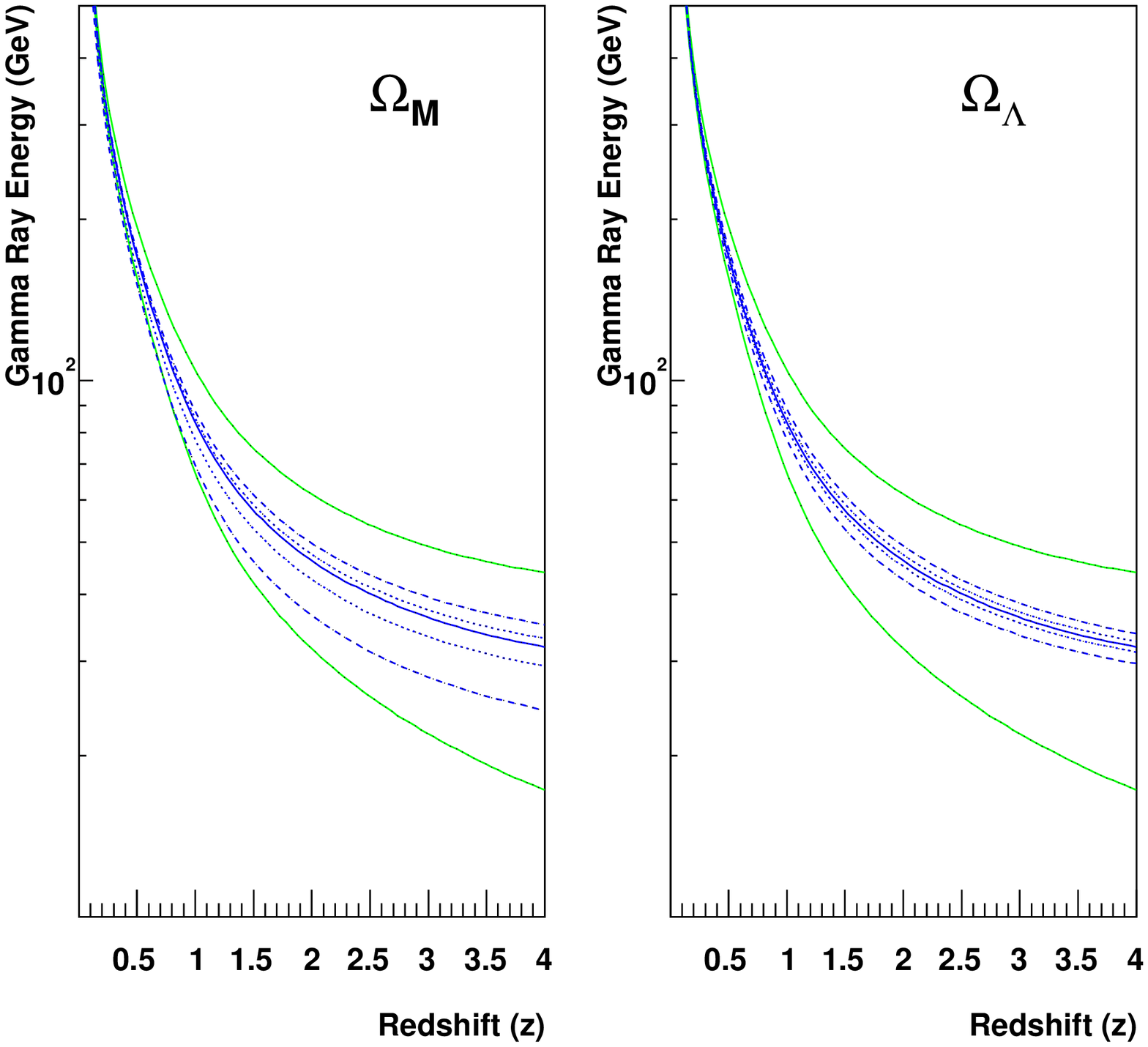,width=8cm}
  \end{center}
    \caption{Gamma Ray Horizon for different values of the
    cosmological densities. In both plots the upper solid line
    is for $\Omega_{M}=1$ and $\Omega_{\Lambda}=0$
    and vice versa for the lower solid line. The dashed lines are
    for $\pm 3 \, \sigma$ and dotted lines are
    for $\pm 1 \, \sigma$ according to the current best fit.}
    \label{fig:Densities}
\end{figure}

\begin{figure}[h]
  \begin{center}
    \epsfig{file=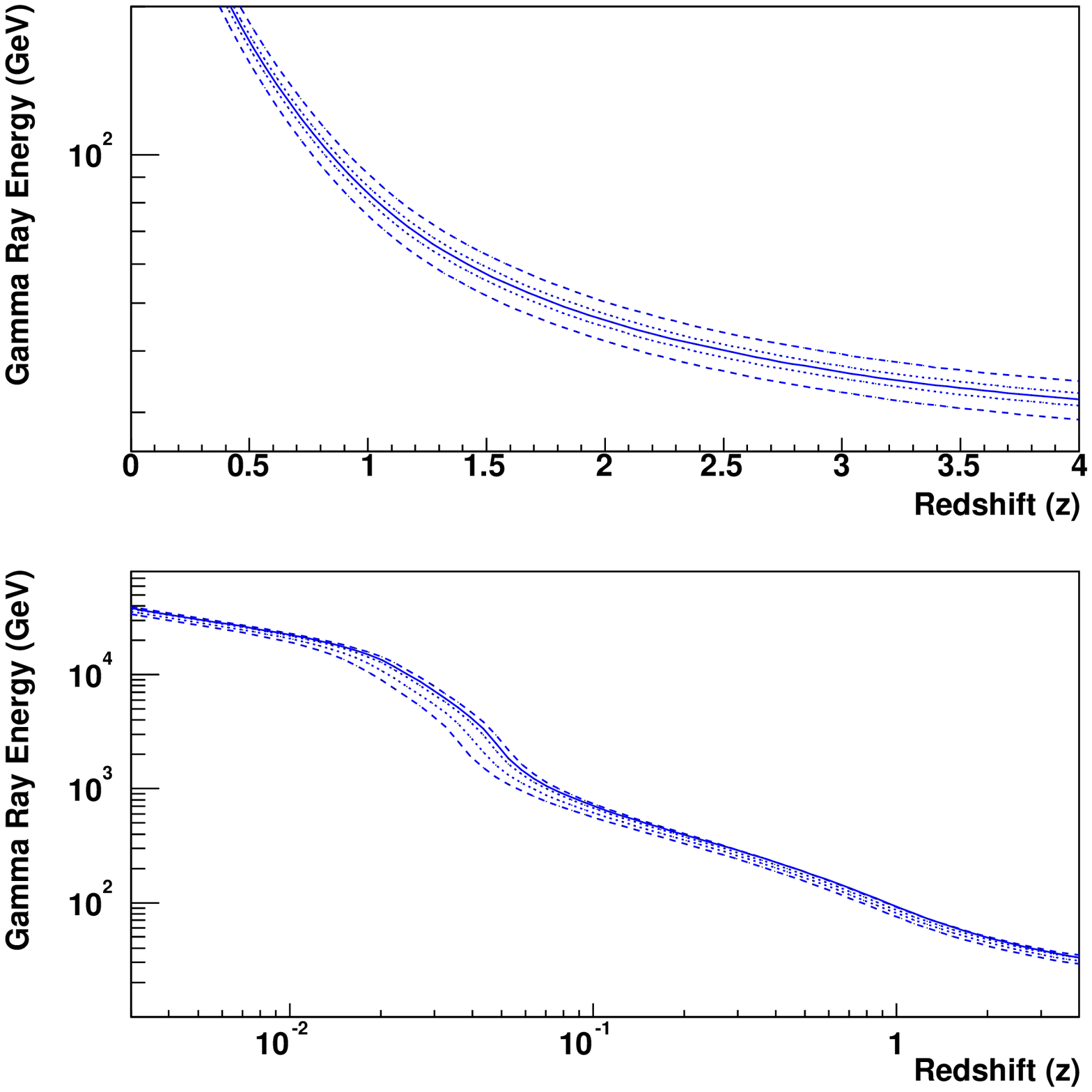,width=8cm}
  \end{center}
    \caption{Gamma Ray Horizon for different values of the Hubble
    constant in linear and logarithmic redshift scales. The
    solid line is for the best fit values. The dashed lines are for
    $\pm 3 \, \sigma$ and dotted lines are for $\pm 1 \, \sigma$ according
    to the current best fit.}
    \label{fig:Hubble}
\end{figure}

For that, first each one of the parameters was changed
$\pm 3 \, \sigma$ from its best fit
value, keeping the rest at their best fit value. The results are shown
in figures \ref{fig:Densities} and \ref{fig:Hubble}. In figure
\ref{fig:Densities},
one can see that a $3 \, \sigma$ variation leads to
a change in the GRH
prediction at high redshift which is of $\sim 8$ GeV for $\Omega_M$
and $\sim 4$ GeV for $\Omega_{\Lambda}$, while
keeping the GHR prediction unchanged at low redshifts as it was expected since
for $z << 1$ the lookback time curve does not depend on $\Omega_{M}$
and $\Omega_{\Lambda}$. Figure~\ref{fig:Hubble} shows that a $3 \, \sigma$
variation on $H_{0}$ also leads to $\sim 5$ GeV
difference at high redshift but there is now also a sizeable difference
for low redshifts, in contrast to the behaviour in the previous case.\par

The Hubble constant enters in the Optical Depth calculation as global
factor and therefore its variation produces a global shift of the
Optical Depth. Then, the flatest zone of the
Optical Depth crosses the $\tau=1$ line at different redshift, which
is seen in the GRH as a region where the logarithm of the GRH energy
as a function of the logarithm of the redshift shows a hard slope
(figure~\ref{fig:Hubble}). Therefore the zone close to the hard slope
region is very sensitive to $H_{0}$, since a 3 $\sigma$ variation
changes $\sim$ $50 \%$ the GRH.\par

\begin{figure}[h]
  \begin{center}
    \epsfig{file=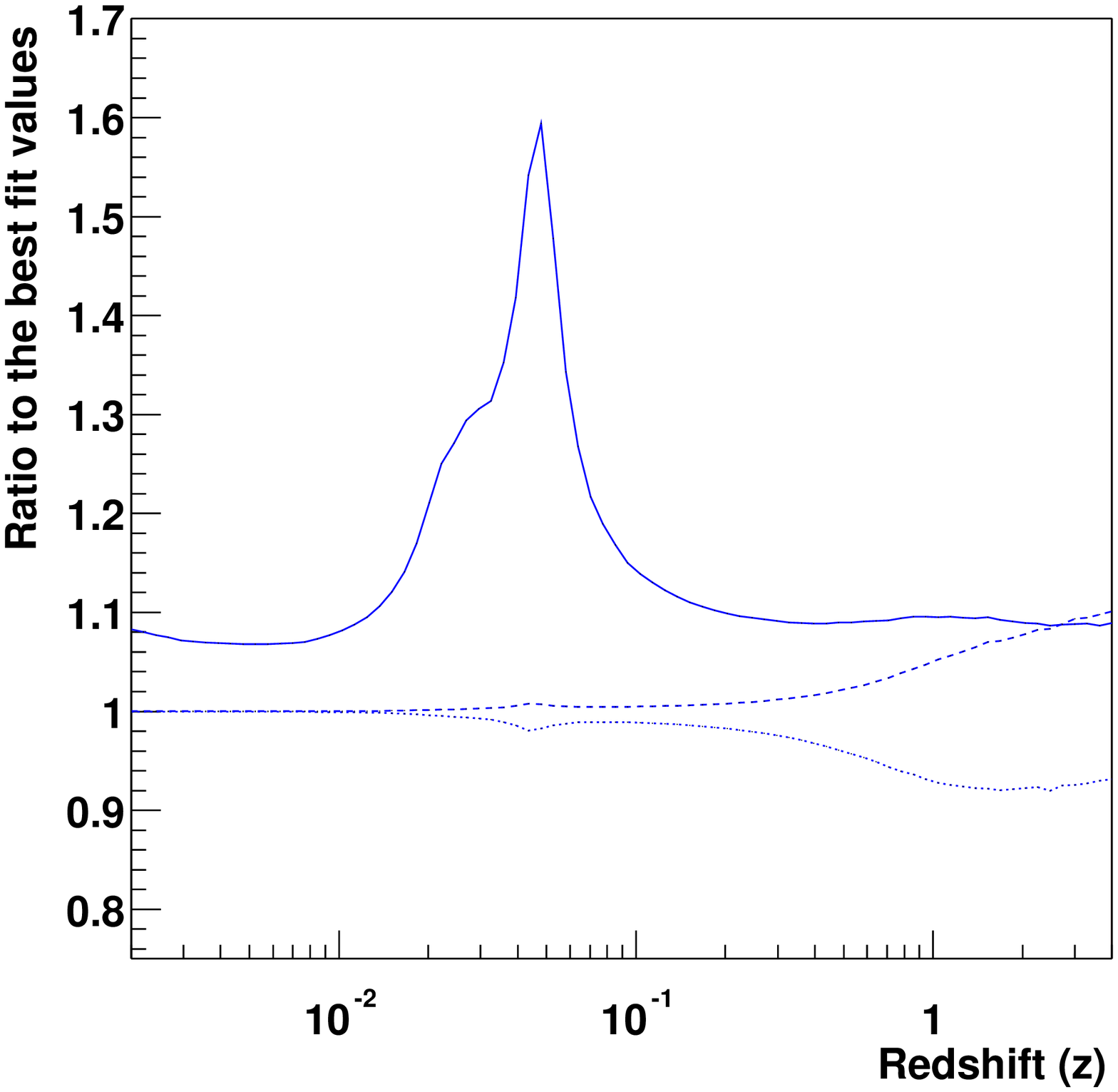,width=8cm}
  \end{center}
    \caption{Ratio of the GRH for values of $H_{0}$
    (solid line), $\Omega_{M}$ (dashed line) and
    $\Omega_{\Lambda}$ (dotted line) which are
    $3 \, \sigma$ above the current best fit, over the GRH for the current
    best fit values }
    \label{fig:Disen}
\end{figure}

The fact that the variations in the GRH due to the Hubble constant and
due to the cosmological densities are qualitatively different leaves
some room to disentangle both kind of parameters. Actually, in
figure~\ref{fig:Disen} it can be seen that a $3 \, \sigma$ difference in
each parameter
produces a change of around $10 \%$ in
both cases at large redshift. But while decreasing redshift the effect
due to changing the cosmological densities goes to zero, the
effect due to the
Hubble constant remains at around $8 \%$. Therefore the precise determination
of the GRH for $z<0.1$ and for $z>0.1$ will allow to perform independent
measurement of both sets of parameters. In fact, above $z>0.1$ also the
dependence on $\Omega_{\Lambda}$ and $\Omega_M$ is different and
therefore, precise measurements may provide a handle to measure
both independently.\par

\begin{figure}[h]
  \begin{center}
    \epsfig{file=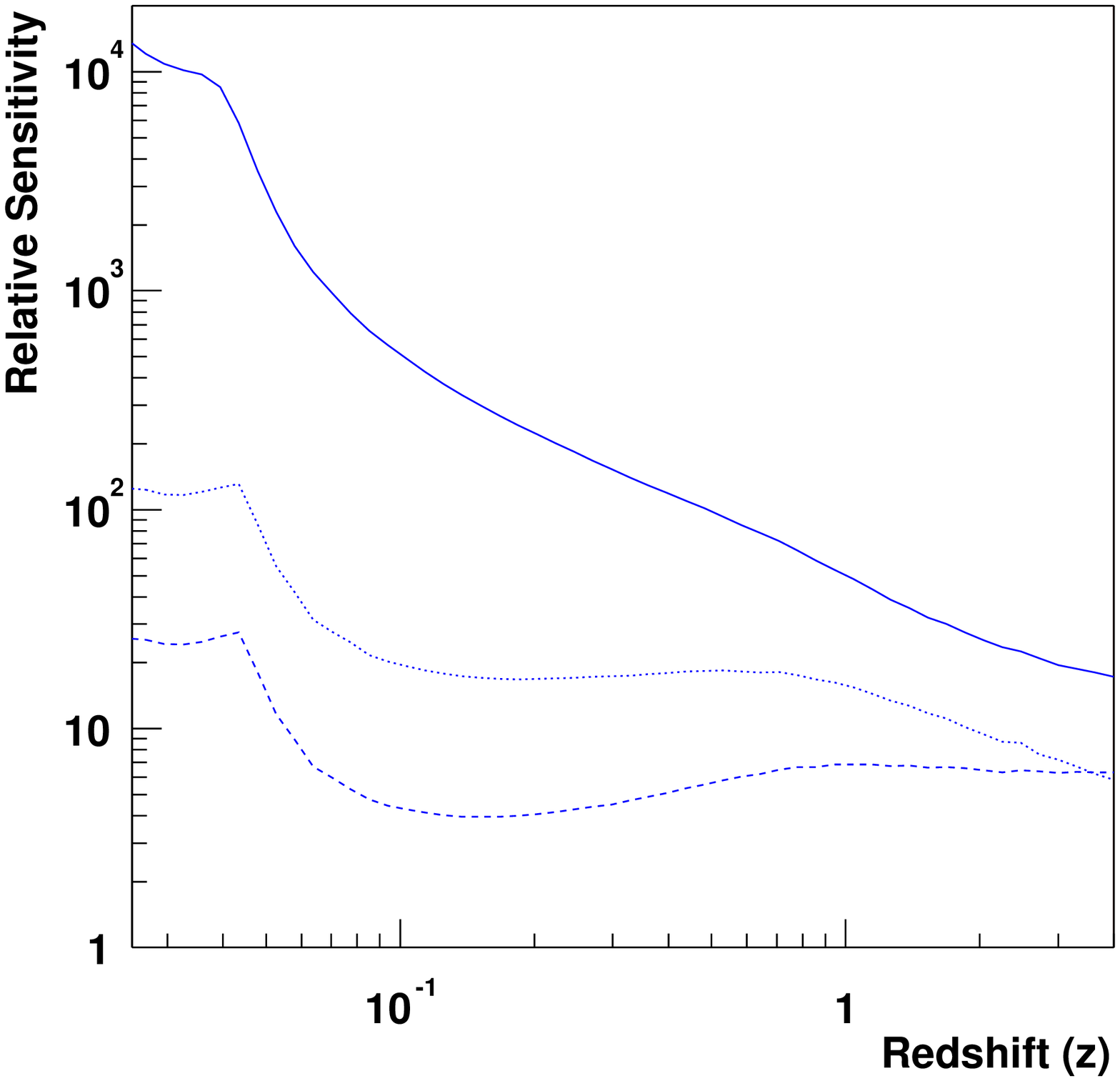,width=8cm}
  \end{center}
    \caption{Sensitivity of the GRH energy to relative variations in $H_{0}$
    (solid line), $\Omega_{M}$ (dashed line) and
    $\Omega_{\Lambda}$ (dotted line).}
    \label{fig:Sensitivity}
\end{figure}

Finally, the sensitivity of the measurement of the GRH energy as a function
of the redshift $z$ on each one of the parameters varied
independently while keeping the rest at their best fit value, has been
computed and is plotted in figure
\ref{fig:Sensitivity}. In that figure the sensitivity for each
parameter $p$ is actually defined as

\begin{equation}
S_p(z) \equiv  p \, \frac{d E_{GRH}(z)}{d p}
\end{equation}

in such a way that for a given uncertainty in the estimation of the
GRH energy $\Delta E_{GRH}$ the relative precision in the
single-parameter determination of $p$ would be

\begin{equation}
\frac{\Delta p}{p} = \frac{1}{S_p} \, \Delta E_{GRH}
\end{equation}

It is clear that the maximal relative sensitivity is for the $H_0$
parameter while for $\Omega_M$ the relative sensitivity is, depending on the
$z$ region, between one and two orders of magnitude smaller and
around a factor 5 even smaller for $\Omega_{\Lambda}$. In this
figure it is also clear that the sensitivities evolve differently with
$z$ and therefore, if precision measurements are obtained it should be
possible to fit simultaneously the three parameters. This possibility
will be explored in detail in the second part of this work already
mentioned at the introduction.\par

\subsection{Beyond the ``standard'' calculation.}

In the calculation presented above, the assumptions taken
are based on our present knowledge of fundamental interactions,
astrophysics and cosmology. Nevertheless, at such high energies
and cosmological distances, for instance the effects from physics
beyond the ``Standard Model'', such as Quantum Gravity or
Supersymmetry, could be important.\par

There are plausible scenarios ``beyond'' the present knowledge
which could affect the GRH prediction and hence, should be
considered. In the following we would
like to comment on our understanding on how these effects, and other
effects not considered in our calculations, could change the GRH
predictions presented above.\par

\subsubsection{The absorption mechanism.}

 So far the only considered absorption mechanism has been the
 gamma-gamma interaction. As we have already seen, the gamma-gamma
 reaction has a strong dependence on the final state fermion mass and
 we have checked explicitly with our calculation that the
 contribution coming from Standard Model fermions other than
the electrons adds a negligible absorption. As far as we know no
extension of the Standard Model provides any alternative light final
 state particle not excluded already by the present accelerators that
 could add any significant amount of gamma-gamma absorption.
 Therefore, no sizable change in the GRH prediction can be expected
in Standard Model
 extensions such a Supersymmetry due to modifications in the
 gamma-gamma cross section.\par

 It is clear that the target for the high energy gammas could also be any
 other particle filling the intergalactic space. Therefore, it
 could be neutrinos, visible matter and barionic and non-barionic dark matter.
 Given the expected density for these targets and the present
 constraints in the dark matter candidates, we are not
 aware of any absorption mechanism with these targets that could add
 any sizable absorption contribution to the one of the gamma-gamma
 reaction for the gamma ray energy range
 considered in this paper and hence, give any sizable correction to the
 GRH prediction.\par

\subsubsection{Lorentz Invariance Violation.}

High energy gamma rays traversing cosmological distances should notice
the quantum fluctuations in the gravitational vacuum which unavoidably
should happen in any
quantum theory of gravitation. These fluctuations may occur on scale
sizes as small as the Planck length $L_P\simeq
10^{-33}$ cm or time-scales of the order of $t_P \simeq
1/E_P$ ($E_P \simeq 10^{19}$ GeV).\par

These gammas will therefore experience a ``vacuum polarization''
correction which should be very small ($O(E/E_{QG})$ where $E$ is
the gamma energy and $E_{QG}$ is an effective scale for Quantum
Gravity, which might be as large as $E_P$ but might become
measurable after the gamma has traversed cosmological distances.
In this Quantum Gravity scenario emerges naturally the requirement
of local "violation" of the Lorenz-Invariance symmetry
\cite{Coleman,Nature} providing as a direct effect an
energy-dependent propagation speed for electromagnetic waves.\par

This local "violation" of the Lorenz-Invariance symmetry changes
the threshold condition for the $\gamma\gamma \rightarrow f^+ f^-$
reaction in a way that depends on the Quantum Gravity model
considered and its effective scale \cite{Amelino-Planck}. For
plausible models, the correction to the GRH predictions turns out
to be quite important and hence, deserves a detailed discussion,
which we presented in reference \cite{LID}.\par

\subsubsection{Astrophysical considerations.}

The gamma-gamma cross section depends strongly on the gamma
polarization state. The calculation made in this paper assumes
unpolarized gammas but it might happen that the specific gamma ray
source producing the high energy gammas under study produces them
with a non-negligible degree of polarization. If that is the
case, the cross section could change in such a way that the GRH
could differ from the above predictions for that specific source.\par

Similarly, in the whole calculation it has been assumed that the
distribution of the EBL was uniform and isotropic at any scale.
Given the fact that we consider cosmological distances this
assumption is quite plausible. Nevertheless, for any specific
gamma ray source, it might happen that the ``local'' EBL density might
differ sizably from the ``average'' one and therefore, the GRH
observed from that source could be sizably different from our prediction.\par

These aspects and other of similar kind depending on the specific
characteristics of the source and its environment should be easy to
disentangle from the fundamental predictions
if enough sources are observed at each redshift location range.\par

\section{Conclusions}

A complete calculation of the Gamma Ray Horizon (GRH) in the gamma ray
energy range which will be covered by the next generation of Gamma Ray
Telescopes has been presented and discussed in detail.\par

Several approaches for the calculation
of the extragalactic background light (EBL) density ranging in
complexity have been compared. That comparison shows that the
uncertainties due to the EBL modeling might be quite large both at low
and high redshift. Nevertheless, the results for the
most realistic approaches agree in predicting that the GRH energy at
large redshifts is of $\sim 30$ GeV and, hence, should be on the reach
of the next generation Cherenkov Telescopes.\par

Following these predictions, the
observable universe should become
transparent to gamma rays of below
$\sim 30$ GeV and then new, high redshift, high energy gamma ray
sources should be observable by the next generation Cherenkov
telescopes
\footnote{A different scenario but with a similar spirit was analysed
in reference \cite{Coppi}. There, the energy threshold for the new
generation gamma-ray detectors was assumed to be $>100$ GeV, well
above the asymptotic horizon given by the GRH predictions discussed
here. Given this fact, to explore the cosmological potential that work
does not use a direct measurement of the GRH but, instead,
the observation of the halo radiation coming from secondary gamma
emission in the electromagnetic cascade generated by the absolved
primary VHE gammas}.\par

If these new sources are abundant enough to make possible a precise
measurement of the GRH energy as a function of the redshift, then either
they can be used to place strong constraints
in the EBL modeling or as a
new technique allowing an independent determination of the
cosmological parameters.\par

Exploring deeper this second scenario, the actual dependence of the
GRH predictions on the
cosmological parameters has been discussed in detail. This
study shows the potential capability of a precise GRH energy
determination as a function of the redshift $z$ to disentangle the
relevant cosmological parameters and provide competitive determinations.\par

A more quantitative study on the actual experimental possibilities
to fit the cosmological parameters with the foreseen observations
is presented in the second part of this work.\par

\section*{Acknowledgements}

We want to thank T.Kneiske and K.Mannheim for many discussions,
for providing us with their EBL spectra and for helping us
to cross-check some preliminary results of this work. We are
indebted to N.Magnussen for his
cooperation in the early stages of this study. We want to thank
our colleagues of the MAGIC collaboration for their comments and
support. We want to thank also G.Goldhaber, P.Nugent and S.Perlmutter
for their encouraging comments.





\end{document}